\documentclass[11pt,twoside]{article}
\usepackage{newpasp}

\usepackage{epsf}

\index{Clarke, T.E.}
\index{Gladders, M.}
\index{Mall\'en-Ornelas, G.}

\begin{document}

\title{Intermediate Velocity Clouds -- Distances, Infall and Shadowing}
\author{T.\ E.\ Clarke}
\affil{NRAO, 1003 Lopezvile Rd., Socorro, NM  87801  USA}
\author{M.\ Gladders}
\affil{Dept.\ of Astronomy, U.\ of Toronto, Toronto, ON, M5S 3H8  Canada}
\author{G. Mall\'en-Ornelas}
\affil{Depto.\ de Astronom\'{\i}a, U.\ Cat\'olica de Chile, Santiago
22, Chile}


\begin{abstract}
We present the current distance estimates to 11 intermediate velocity
clouds (IVCs) which are part of the David Dunlap Observatory
Intermediate Velocity Cloud (DDO IVC) Distance Project (Clarke et al.\
1999). We briefly discuss constraints these place on infall models and
show the potential of IVCs to constrain the radial distribution of hot
gas in the Galaxy through shadowing of the soft X-ray background.
\end{abstract}




\section{Distances and Infall}

For each IVC, a set of target stars was observed spectroscopically in
the Na\,{\sc i} doublet region for evidence of absorption from the
target IVC (details in Gladders et al.\ 1998). The distances and
derived cloud masses are presented in Table~1.

The IVC distances permit us to compare cloud velocities with Galactic
infall models. The general trend of the estimated infall velocities
appears consistent with an increase in velocity with increasing
height; this trend is not expected for clouds following ballistic
trajectories (see Clarke et al.\ 1999 for details).

\noindent

\begin{table}[h]
\centering
\tiny{
\begin{tabular}{cllr} \hline
{\bf Cloud(s)} & {\bf ${\rm V_{LSR}}$} & {\bf\ \ \ \
\ \ d (pc)} & {\bf $M_{HI}$ ($M_{\odot}$)} \\ \hline 
G163.9+59.7 & -19 & ${\rm 543^{+514}_{-243}}$ -- ${\rm 1533^{+844}_{-541}}$
& 1600 -- 12700\\
G139.6+47.6 \& G141.1+48.0 & -12.5 & ${\rm 233^{+92}_{-116}}$ -- ${\rm
325^{+93}_{-66}}$ & 30 -- 150 \\
G135.5+51.3 & -47.2 & ${\rm 359^{+157}_{-109}}$ -- ${\rm
1970^{+1012}_{-659}}$ & 860 -- 25600 \\
G149.9+67.4 & -6.3 & ${\rm 246^{+108}_{-58}}$ -- ${\rm
550^{+150}_{-150}}$ & 270 -- 1370 \\
G249.0+73.7 & -0.6 &  $<$ ${\rm 550^{+375}_{-200}}$ & $<$ 3170 \\
G124.1+71.6 & -11.4 & $<$ ${\rm 267^{+104}_{-75}}$ & $<$ 170 \\
G107.4+70.9 \& G99.3+68.0 & -28.3 & ${\rm 742^{+324}_{-217}}$ -- ${\rm
823^{+393}_{-251}}$ & 2700 -- 3300 \\
G86.5+59.6 & -39 & $>$ ${\rm 1140^{+560}_{-375}}$ & $>$ 9600 \\
G90.0+38.8 \& G94.8+37.6 & -23.6 & ${\rm 463^{+192}_{-136}}$ -- ${\rm
618^{+243}_{-174}}$ & 600 -- 1080 \\ 
G81.2+39.2  &  +3.5 & ${\rm 450^{+187}_{-132}}$ -- ${\rm 851^{+407}_{-276}}$
& 1180 -- 4250 \\
G86.0+38.3 & -43.4 & ${\rm 921^{+443}_{-280}}$ -- ${\rm
2366^{+1192}_{-793}}$ & 450 -- 3000 \\ \hline 
\end{tabular} }
\caption{\small The current distance estimates for the sample of 11
clouds in the DDO IVC project (Draco is G90.0+38.8 \& G94.8+37.6). The
masses were determined by integrating the H\,{\sc i} column
density over the extent of the cloud.}
\end{table}

\section{Shadowing}

The nature of the SXRB is in dispute.  One model suggests that most of
the SXRB arises locally from the low density plasma surrounding the
Sun and extending for a few hundred parsecs (Sanders et al.\
1977). Other models propose that most of the SXRB originates
at large distances in a Galactic halo (e.g.\ Marshall and Clark
1984).  

Soft X-ray background (SXRB) measurements toward Draco show a
reduction of $\sim$ 50\,\% in the 1/4 keV emission relative to the
adjacent sky, consistent with photoelectric absorption of the SXRB by
Draco (Burrows and Mendenhall 1991; Snowden et al.\
1991). Furthermore, the profile of the cloud's dust emission
(Figure~1a) is closely followed by the X-ray shadowing (Figure~1b).
Our distance to Draco places it beyond the local plasma and therefore
the X-ray emission it is shadowing must have an origin at larger
Galactic distances.  Specifically, 50\,\% or more of the SXRB emission
originates from distances greater than 460 pc.  Note that a sample of
X-ray shadowing clouds at known distances (such as the DDO IVC sample)
can be used to infer the detailed distribution of the SXRB through
determining the foreground and background components (work in progress).

\vspace{-0.5cm}
\begin{figure}[!h]
\plottwo{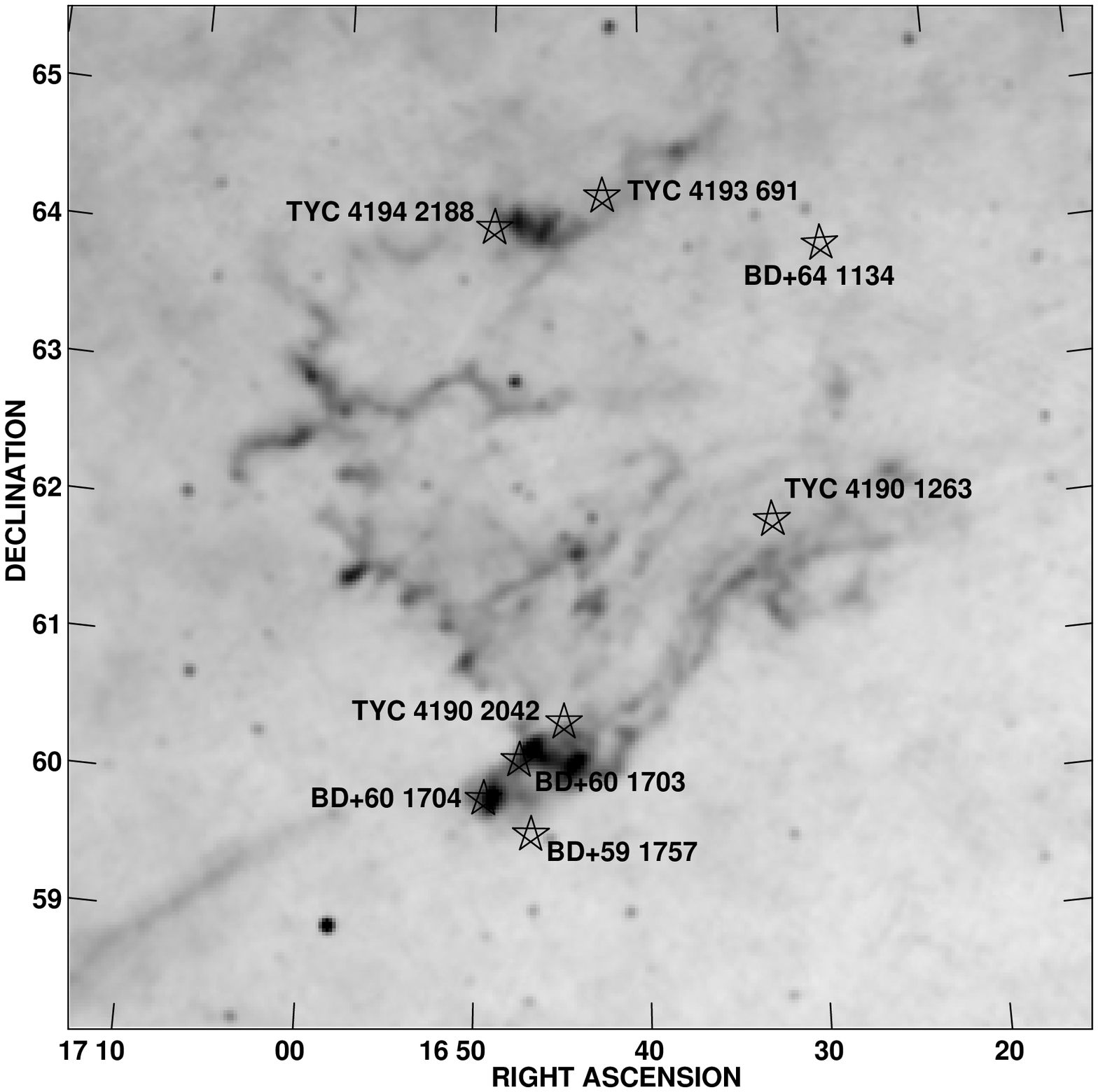}{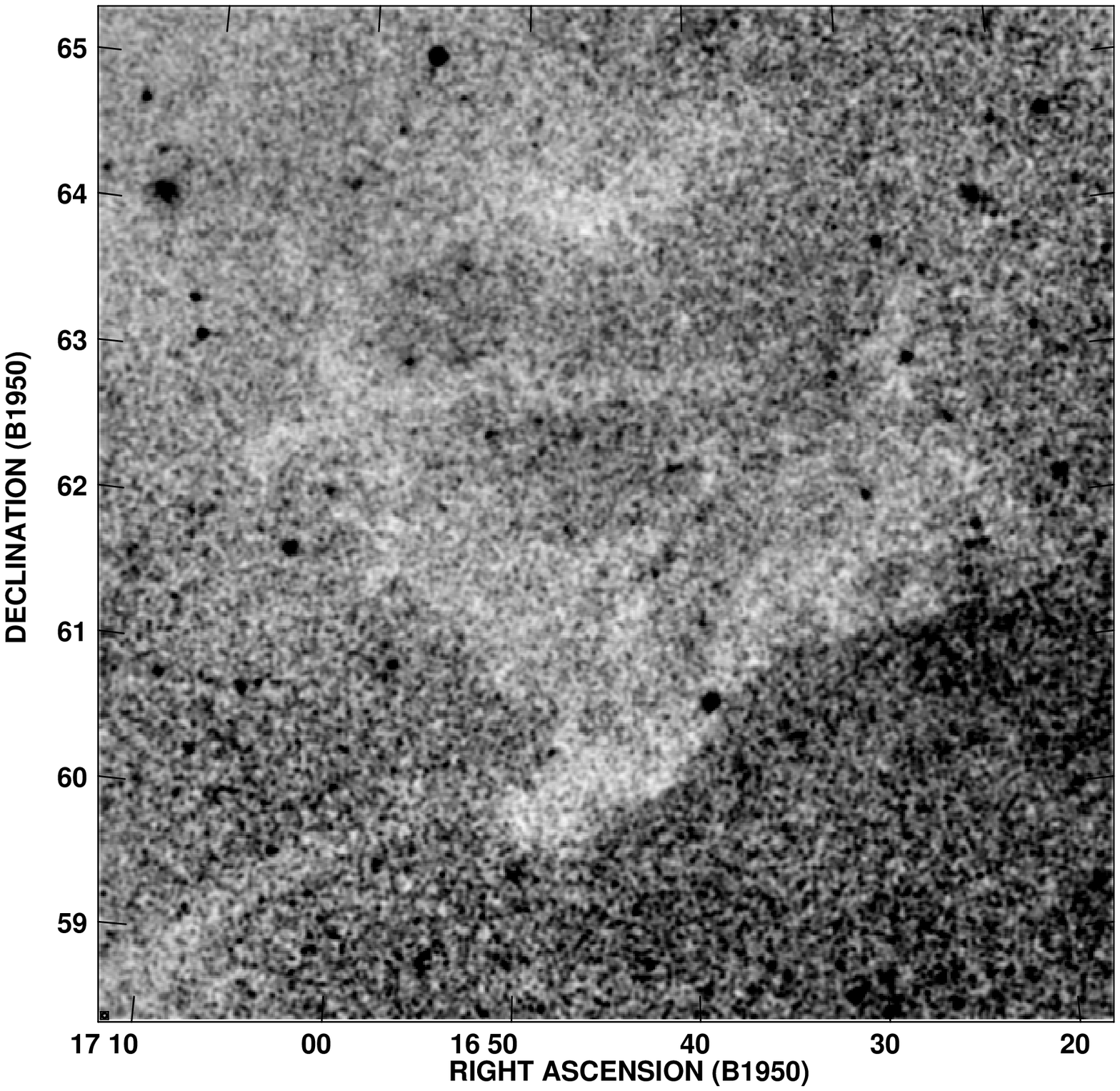}
\caption{\small Left: the IRAS 100 $\mu$m image of the Draco Cloud
with the positions of eight program stars indicated. Right: the ROSAT
All Sky Survey 1/4 keV X-ray emission in the direction of the Draco
Cloud. Notice the shadowing of the X-ray emission with the shape of
the cloud.}
\label{fig1}
\end{figure}


\acknowledgements
{\small We acknowledge the work of the entire DDO IVC Distance Project team
and thank the organizers of this meeting.}

\smallskip
\noindent
{\bf Refernces}

\noindent
{\small Burrows, D.N., \& Mendenhall, J.A., Nature, 351, 629\\
Clarke, T.E., et al.\ 1999, ASP Conference Series 168\\
Gladders, M.D. et al., 1998, ApJ, 507, 161\\
Marshall, F.J., \& Clark, G.W.\ 1984, ApJ, 287, 633\\ 
Sanders, W.T., et al.\ 1977, ApJL, 217, 87\\
Snowden, S.L., et al.\ 1991, Science, 525, 1529\\
}
\end{document}